\pdfoutput=1
\documentclass[sigconf]{acmart}

\usepackage{booktabs} 
\usepackage{listings}
\usepackage{tabularx}
\usepackage{color}
\usepackage{colortbl}
\usepackage{enumitem}
\usepackage[ruled,lined,linesnumbered,vlined,algo2e]{algorithm2e}

\newcommand{\etc}{\mbox{\emph{etc.\ }}}
\newcommand{\eg}{\mbox{\emph{e.g.\ }}}

\copyrightyear{2018} 
\acmYear{2018} 
\setcopyright{acmcopyright}
\acmConference[ICSE-SEIP '18]{40th International Conference on Software Engineering: Software Engineering in Practice Track}{May 27-June 3, 2018}{Gothenburg, Sweden}
\acmPrice{15.00}
\acmDOI{10.1145/3183519.3183554}
\acmISBN{978-1-4503-5659-6/18/05}

\begin{document}
\title[SmartUnit: Empirical Evaluations for Automated Unit Testing\\ of Embedded Software in Industry]{SmartUnit: Empirical Evaluations for Automated Unit Testing of Embedded Software in Industry}

\author{Chengyu Zhang\textsuperscript{1}, Yichen Yan\textsuperscript{1}, Hanru Zhou\textsuperscript{1}, Yinbo Yao\textsuperscript{2}}
\author{Ke Wu\textsuperscript{2}, Ting Su\textsuperscript{3$*$\authornote{Geguang Pu, Weikai Miao and Ting Su are the corresponding authors.}}, Weikai Miao\textsuperscript{1$*$}, Geguang Pu\textsuperscript{1$*$}}
\renewcommand{\authors}{Chengyu Zhang, Yichen Yan, Hanru Zhou, Yinbo Yao, Ke Wu, Ting Su, Weikai Miao, Geguang Pu}
\affiliation{\textsuperscript{1}School of Computer Science and Software Engineering, East China Normal University, China} 
\affiliation{\textsuperscript{2}National Trusted Embedded Software Engineering Technology Research Center, China}
\affiliation{\textsuperscript{3}School of Computer Science and Engineering, Nanyang Technological University, Singapore}
\email{dale.chengyu.zhang@gmail.com, sei_yichen@outlook.com, hanruzh@gmail.com, snowingsea@gmail.com}
\email{bukawu@126.com, suting@ntu.edu.sg, wkmiao@sei.ecnu.edu.cn, ggpu@sei.ecnu.edu.cn}

\renewcommand{\shortauthors}{C.~Zhang, Y.~Yan, H.~Zhou, Y.~Yao, K.~Wu, T.~Su, W.~Miao, G.~Pu}

\begin{abstract}
In this paper, we aim at the automated unit coverage-based testing for embedded software.  To achieve the goal,  by analyzing the industrial requirements and our previous work on automated unit testing tool CAUT,  we rebuild a new tool, SmartUnit,  to solve the engineering requirements that take place in our partner companies. 
SmartUnit is a dynamic symbolic execution implementation, which supports statement, branch, boundary value and MC/DC coverage. 

SmartUnit has been used to test more than one million lines of code in real projects. For confidentiality motives, we select three in-house real projects for the empirical evaluations. We also carry out our evaluations on two open source database projects, SQLite and PostgreSQL, to test the scalability of our tool since the scale of the embedded software project is mostly not large, 5K-50K lines of code on average. From our experimental results,   in general, more than 90\% of functions in commercial embedded software achieve 100\% statement, branch and MC/DC coverage, more than 80\% of functions in SQLite  and more than 60\% of functions in PostgreSQL achieve 100\%  statement and branch  coverage. Moreover, SmartUnit is able to find the runtime exceptions at the unit testing level. We also have reported exceptions like \textit{array index out of bounds} and \textit{divided-by-zero} in SQLite. Furthermore, we analyze the reasons of low coverage in automated unit testing in our setting and give a survey on the situation of manual unit testing with respect to automated unit testing in industry.

\end{abstract}



\maketitle

\section{Introduction}

Embedded software widely exists in various control systems, which is mostly specialized for the particular hardware it runs on and have different constraints, like time or memory. Manufacturers have broadly developed all sorts of embedded software in the electronics, \eg cellphones, robots, digital TV  \etc.  Moreover,  most of the equipments in industrial infrastructure extensively use embedded software, for instance, control systems in cars,  trains,  power plants, satellites, and so on.  Thus, how to ensure the reliability and dependability of embedded software is an ongoing challenge for the safety-critical embedded systems.

Software testing is one of the most common ways to ensure the software quality. 
Many developers and researchers concentrate on how to improve the effectiveness and efficiency of the testing methods to achieve higher coverage and find more faults. 
Unit testing is  an important step to ensure the software quality during the stage of software development  ~\cite{ieee1990standard, runeson2006survey}.
For example, 79\% of Microsoft developers use unit testing in their daily work~\cite{venolia2005software}. Meanwhile,  unit testing is a mandatory task required in various international standards for different industrial systems, \eg, IEC 61508, ISO26262, RTCA DO-178B/C \etc.  For instance,  IEC61508, which is intended to be designed as a basic functional safety standard applicable to all kinds of industry specifications, such as Safety Integrity Level (SIL),  to provide a target to attain with respect to a system's development.  If the software is up to level SIL 3/SIL4,  both branch and MC/DC coverages have to be achieved to 100\%  during the unit testing stage.  If not achieved, engineers are required to explore the software codes and find the reasons.

In general, the main objective in unit testing is twofold. One is to verify that the functionality is correct at the function level and the other is to ensure the function is fully tested and all possible branches and paths are taken. We call the former \textit{functionality testing}  and the latter \textit{coverage-based testing}.  Functionality testing is carried out in almost every software company as the basic quality assurance means. Software engineers design the test cases manually in regards to the software design specification and then run the test cases to check the final results by specification or assertion.  For coverage-based testing, software engineers may go through the codes and compute the conditions on branches/paths to obtain the test cases.   During those activities, they usually utilize the commercial unit testing tools like Testbed\footnote{http://ldra.com/industrial-energy/products/ldra-testbed-tbvision/}, VectorCAST\footnote{https://www.vectorcast.com/} \etc, to help them accelerate the task of test data design.  Whatever functionality testing or coverage-based testing is involved,  the work of test data design almost depends on manpower, which is a tedious job for software engineers. 

In this paper, we aim at the unit coverage-based testing, and we believe that with the great advance achieved in the field of automated testing~\cite{shamshiri2015automatically, fraser2014large, almasi2017industrial}, especially in symbolic execution~\cite{cadar2008klee,sen2005cute, godefroid2005dart, su2014automated} and decision procedure~\cite{barrett2013decision}, we are capable of fully automatizing coverage-based testing in order to largely save manpower.  To achieve this goal,  firstly, we have elaborately investigated the real unit testing requirements from selected ten partner companies in China Mainland, covering main safety-critical fields like railway, aerospace, nuclear plant, and automobile. Secondly,  by analyzing the collected requirements and based on our previous work on automated unit testing tool CAUT~\cite{su2014automated, su2015combining,SuPMHS16},  we rebuild a new tool, \textit{SmartUnit},  to meet the real engineering requirements that take place in those companies. 

We observed that most of the companies bought kinds of commercial unit testing tools like Testbed or Tessy\footnote{https://www.razorcat.com/en/product-tessy.html}, which can support different chip platforms. They totally design test data by hand. In these ten companies, no one has used the automated testing tool in their production departments, but two of them have tried test data generation tools. The main reason for not adopting them is that existing commercial automated testing tools have achieved very low coverage but large test suites, since most of the existing tools are based on the random testing technology or simple branch analysis while ignoring the path analysis. We will discuss this more in section 5.  For the tools from academia, like KLEE~\cite{cadar2008klee} or  Otter~\cite{ma2011directed} are far from mature in industry.  In a word, it is quite surprising that NONE of the visited ten companies has adopted tools to help test data design, they still use the most traditional approach, manpower,  to test the design for the safety-critical systems while the symbolic execution technique has already achieved great success in other fields like security and verification.  

\textit{SmartUnit} still follows the principle of symbolic execution approach~\cite{king1976symbolic} but has its novelty in the following points especially in the aspect of practice engineering.  

\begin{enumerate}[]
\item It is a dynamic symbolic execution (DSE) implementation. Based on the experience of developing CAUT~\cite{su2014automated}, we elaborately design the execution engine of \textit{SmartUnit} and make it robust enough since, in practice, an embedded project involving  20K lines of code on average cannot stop abnormally.  We also design a new heuristic search strategy for speeding up automated unit testing, which supports statement, branch, boundary value, and MC/DC coverage.

\item It can generate all the stubs automatically. One tedious work for unit testing is to design stubs to replace the existing function calls or global variables etc., the same in using commercial tools as well. SmartUnit makes it simpler. It also provides the options to leave the decisions for software engineers in case of adapting different application scenes. 

\item It is deployed as a private cloud-platform.  Since the symbolic execution engine consumes computing resources heavily, it is designed as a private cloud-platform for internal use. Users only need to update the software project package by the web browser to the server, which will make the whole analysis automatically including stubs generation, symbolic execution, test data report generation \etc.

\item It can be seamlessly integrated into the existent development environment, especially  connecting to these commercial unit testing tools.  Developers get used to the tools at hand,  so one of the design philosophies for SmartUnit is to make the existent testing process as short as possible. To this end, SmartUnit can generate the test data input files for commercial tools like Testbed and Tessy. Once the test data suite is generated, it can be used by users directly in their unit testing tool at hand.
\end{enumerate}

SmartUnit has been successfully applied in our four partner companies at the first stage from May to September 2017.   
For instance,  our partners include \emph{China Academy of Space Technology},  which is the main spacecraft production agency in China (like NASA in the United States); \emph{CASCO Signal Ltd.}, which is the best railway signal corporation in China; 
and \emph{Guangzhou Automobile Group},  which is one of biggest car manufacturers in China. 

SmartUnit has been used to test over one million lines of code in real projects.  For confidentiality motives, we select three in-house projects for the empirical evaluations, but we still cannot present the code example for the same reason. Thus, we carry out our evaluations on two open source database projects \textit{SQLite} and \textit{PostgreSQL}.  We did not select open-source embedded software because we would like to test the scalability of our tool since the  scale of the embedded software project is mostly not large, around 5K-50K lines of code.  On the other hand, database projects are more complex than embedded software. From our experimental results,   in general, more than 90\% of functions in commercial embedded software achieve 100\% statement, branch and MC/DC coverage, more than 80\% of functions in SQLite  and more than 60\% of functions in PostgreSQL achieve 100\%  statement and branch  coverage.  Moreover, SmartUnit has the ability to find the runtime exceptions at the unit testing level. We have also reported exceptions like \textit{array index out of bounds} and  \textit{divided-by-zero} in SQLite.

The organization of the paper is as follows.  Section 2 introduces the  background of our tools and techniques, 
Section 3 shows the overview and the implementation details of our DSE-based C program unit test generation framework and its private cloud-platform, 
Sections 4 and 5 set up our evaluation and analyze the results, 
Section 6 discusses some related work, and Section 7 gives the conclusion.

\section{background}
\subsection{Industry Situation}
Unit testing is an important engineering activity to ensure the quality of software in industry, especially for the manufacturers of safety-critical systems, \eg, the aerospace and railway signal control companies.
Although unit testing is a compulsory engineering activity requested by the standards, its application in industry is still suffering from low coverage and low efficiency due to the lack of automated tool. 
In most cases, test cases are first manually generated by engineers and then executed on the program code by some commercial third-party tools (\eg Testbed, Tessy) to run the program code. Since manual test generation is time-consuming, the companies usually spend a lot of costs to employ test engineers or outsource in producing the test case. According to our industrial partners' experience, a trained test engineer can produce test case for 5-8 functions per day. One of our industrial partner spend over \$10,000 per month for hiring a group of unit testing engineers while still suffers from the low efficiency and low fault detection. To tackle these challenges, a powerful tool that can automatically derive test cases of high coverage is highly desirable. Further, such a tool needs to be seamlessly integrated with mainstream third-party test execution tools. That is, the generated test data can be recognized by these tools to perform testing. It is inefficient and expensive, so they need efficient automatic tools to generate the test case for third-party tools.

\subsection{Dynamic Symbolic Execution}
Symbolic execution was first proposed by James C. King~\cite{king1976symbolic} in 1976.
Due to the limited computing resource and SMT constraint solver, symbolic execution was not a practical technique in those years. 
Thanks to the recent computing resource improvement and a series of fantastic SMT solvers, 
such as Z3~\cite{de2008z3}, STP~\cite{ganesh2007decision, cadar2008exe}, CVC4~\cite{barrett2011cvc4}, \etc,
many symbolic execution engines have come into existence~\cite{cadar2011symbolic}
(\eg  KLEE~\cite{cadar2008klee, cadar2008exe}, DART~\cite{godefroid2005dart}, CAUT~\cite{su2014automated} for C, JPF-SE~\cite{anand2007jpf} for Java).
Researchers have also applied symbolic execution to software testing, including automatically generating test cases~\cite{su2014automated, li2011klover}.

Symbolic execution uses symbolic values as programs inputs to simulate the execution of programs. When dealing with a control-flow fork, symbolic execution engine collects the conditional expressions along the path as path constraint. When reaching the terminal of the program, SMT solver solves all the path constraint to get a result. The result is a test case that follows the path. The symbolic execution stops when all program paths are explored.

\lstset{
    language=C, 
    numbers=left, 
    xleftmargin=2em, 
    xrightmargin=-2em, 
    numberstyle=\scriptsize, 
    escapeinside={@}{@}, 
    basicstyle=\footnotesize\ttfamily, 
    numbersep=8pt, 
    commentstyle=\itshape\scriptsize
}

\begin{figure}
  \begin{minipage}[b]{.4\linewidth}
\begin{lstlisting}
int checkSign(int x){
    if (x > 0)
        return 1;
    else if (x == 0)
        return 0;
    else 
        return -1;
}
\end{lstlisting}%
  \end{minipage}
  \begin{minipage}[b]{.51\linewidth}
\includegraphics[width=\textwidth]{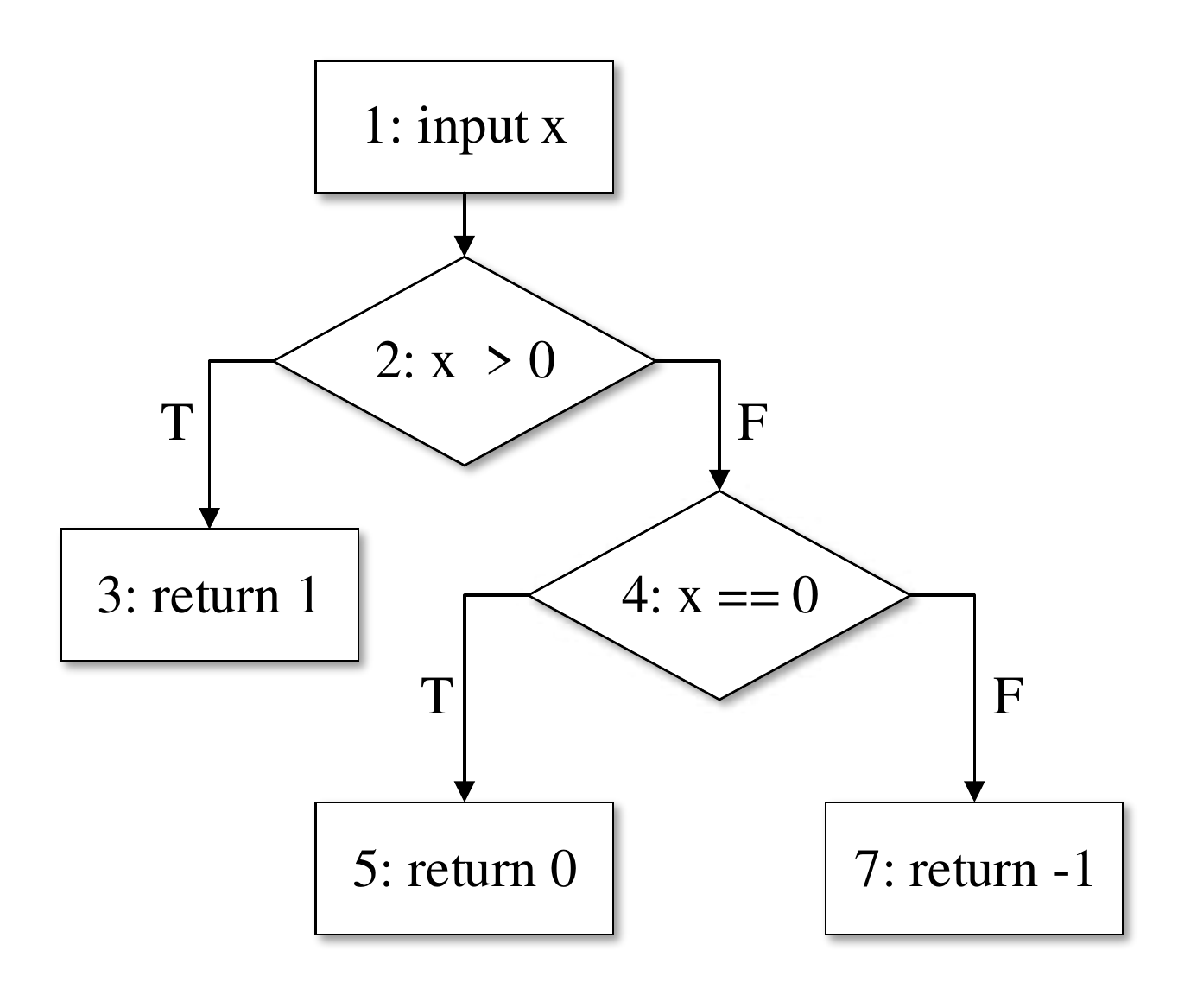}
\label{fig:example_right}
  \end{minipage}
  \caption{An example: \emph{checkSign} function.}
\label{fig:example}
\end{figure}

\textbf{D}ynamic \textbf{S}ymbolic \textbf{E}xecutive (DSE) is a variant symbolic execution,
which was proposed in 2005~\cite{godefroid2005dart, sen2005cute}, also called \textit{Concolic Execution}. 
DSE uses concrete randomly generated values as input to execute the program while collecting path constraints during the execution. Then SMT solver solves a variant of the conjunction of these symbolic constraints to output a new input value. The new input value will be used to execute a new program path.

Figure~\ref{fig:example} is an illustrative example to explain the symbolic execution.  
The code is a function named \texttt{checkSign} for checking signals, with its control-flow graph. If the input variable x is a positive number, the function returns value 1; if x is a negative number, the function returns value -1; otherwise 0;.
The right side part of Figure~\ref{fig:example} is the control-flow graph of the function.

\begin{figure}
    \includegraphics[width=0.5\textwidth]{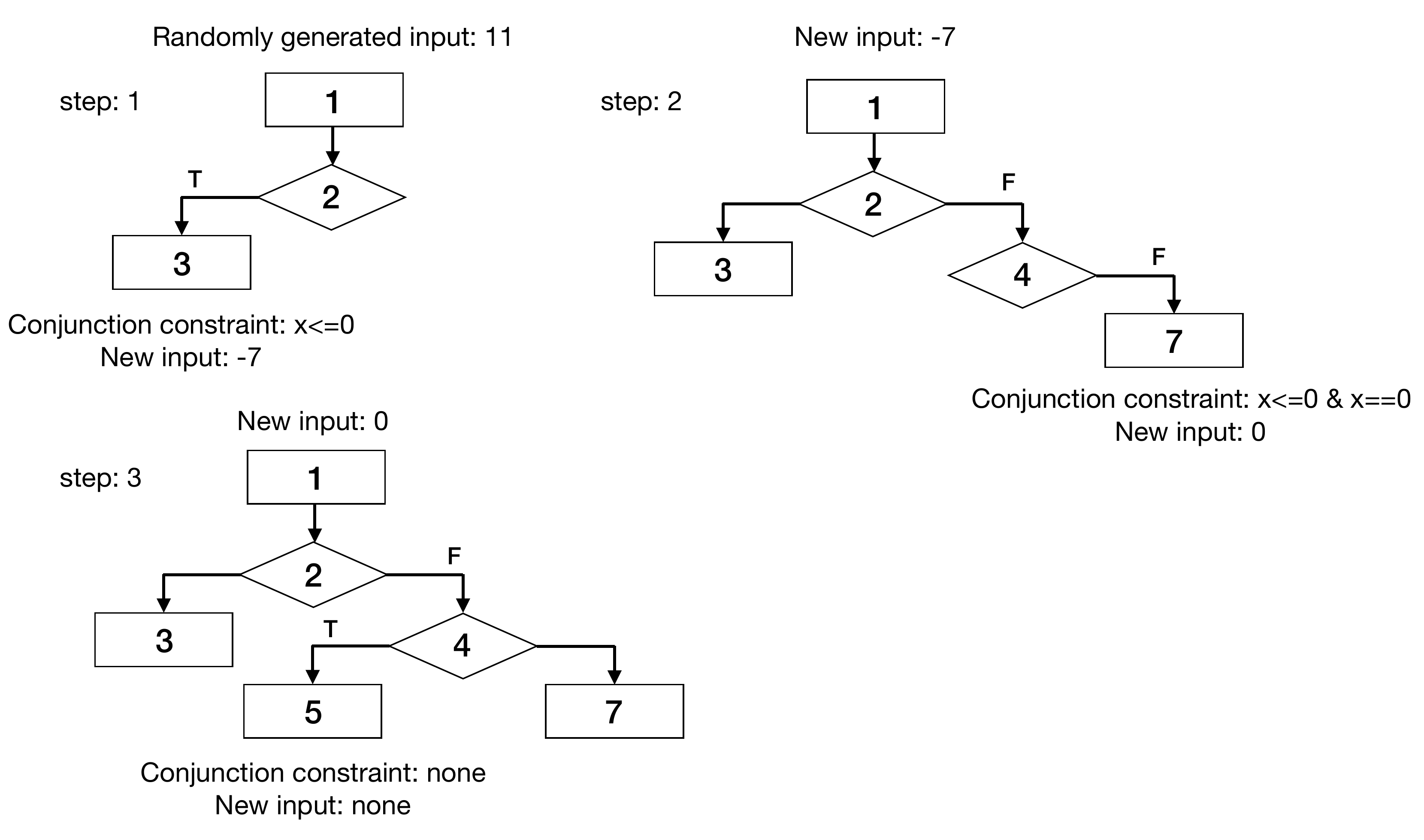}
    \caption{Process of dynamic symbolic execution.}
    \label{fig:DSE}
\end{figure}

Figure~\ref{fig:DSE} describes the process of adopting the DSE. The goal is to cover line 7. 
DSE engine first randomly generates an input value, \eg 11. Using this input value, the program reaches line 3. Then DSE engine negates the constraint collected from line 2 to solve a new input value, \eg -7. Therefore the statement of line 7 is triggered. The DSE engine further negates the conjunction constraint collected from line 2 and line 4, generating another new input value, \eg 0. Ultimately DSE engine can use value 0 as input to reach the statement of line 5.
 
Recently, a variety of DSE-based tools have been proposed~\cite{burnim2008heuristics, cadar2008klee, cadar2008exe, godefroid2005dart, godefroid2008automated, majumdar2009reducing, sen2005cute, tillmann2008pex}.
There are still some challenges for the DSE, \eg exponential growth paths, symbolic pointer, guided execution, \etc. Section 3 will describe the implementation of our SmartUnit DSE-based engine and explain how these problems are solved.

\subsection{Coverage Criteria}
SmartUnit is a coverage-driven unit testing tool. One of its major goals is to generate the test suite towards a high coverage of code. In this subsection, we will introduce some commonly used coverage criteria in the industry.
\subsubsection{Statement Coverage}
Statement coverage requires all the statements in the program code under test be executed at least once by the test cases. Such coverage is easy to measure and the 100\% statement coverage is also easy to achieve. Statement coverage is the most common used coverage criterion. For example, we can use it to detect the statements that are never executed. Since a large number of faults may not be detected by the criterion, it is usually not used alone. 
\subsubsection{Branch Coverage}
Branch coverage is a stronger coverage criterion than statement coverage~\cite{zhu1997software}. 
It needs to confirm all of the possible branches from each decision are executed at least once. The branch coverage is also easy to achieve the 100\% coverage.
\subsubsection{MC/DC coverage}
MC/DC is the abbreviation of Modified Condition/Decision Coverage. It is a stronger coverage criterion than the branch coverage. In general, each decision is an atomic condition or combined with more than one atomic condition. When test cases satisfy each decision with value true and false, it obviously achieves branch coverage. 
However, MC/DC coverage further demands that test cases satisfying an atomic condition should affect decision independently with truth-value true and false, it is called Modified Condition/Decision Coverage. In practice, MC/DC-based unit test is usually difficult for the test engineer to write manually because of the complex logic in decision condition. But MC/DC coverage criterion is required in a variety of industrial standards. Therefore, test engineers always spend a huge amount of costs and time in designing the MC/DC test cases manually.

\lstset{
    language=make, 
    numbers=left, 
    xleftmargin=2em, 
    xrightmargin=-2em, 
    numberstyle=\scriptsize, 
    escapeinside={@}{@}, 
    basicstyle=\scriptsize, 
    numbersep=8pt, 
    commentstyle=\scriptsize
}
\section{Framework Implementation}
\subsection{Framework Architecture}
\begin{figure}
    \includegraphics[width=0.3\textwidth]{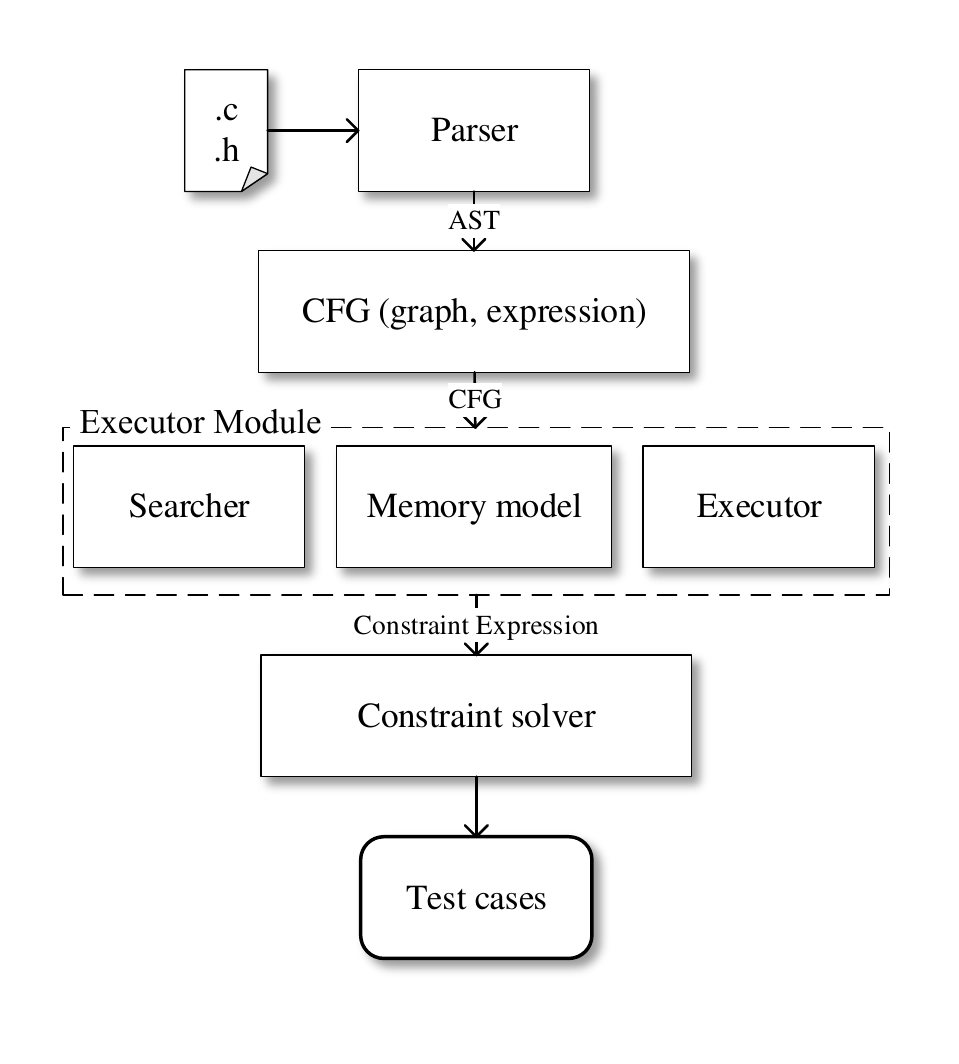}
    \caption{The architecture of SmartUnit.}
    \label{fig:backend}
\end{figure}

Figure~\ref{fig:backend} shows the core framework of SmartUnit.
The basic process of generating test cases in our approach is as follows:

SmartUnit accepts the \texttt{.c} and \texttt{.h} files as its input. To deal with macros and make sure some external symbols can be introduced into the source file, we use \textit{libclang}\footnote{http://clang.llvm.org/doxygen/group\_\_CINDEX.html} as preprocessor to generate processed .c file. Then libclang is also used to parse the processed file to generate the AST (abstract syntax tree).

We establish the CFG (Control Flow Graph) model based on the abstract syntax tree generated in the previous step. It consists of the control flow graphs generated from the proceeded files and the information of variables, expressions, functions \etc. Each node in the control flow graph represents a statement block in the source code. The sequential node contains exactly one incoming edge and one outgoing edge. The branch node contains one incoming edge and more than one outgoing edge and indicates the condition of the branch. The branch node usually represents if-else statement, while statement and switch statement \etc.

\subsection{Executor Module}
Executor module consists of the memory model, executor, and searcher. This part mainly processes the CFG model given by the previous steps.

\subsubsection{Executor}
Executor executes the statement expressions in the current node and drives the searcher to select next edge to explore. The executor does not really execute the C statements, it actually transforms the C statement into blocks, declarations or expressions structures that are stored inside SmartUnit. The executor updates information in the CFG model and adds the constraints to a path, after gathering them from the node statement. When the executor reaches the end node, it collects all the constraints on the path, and solves them by the constraint solver such as Z3.

\subsubsection{Searcher and Search Strategy}
To perform the search on control flow graph, we propose a new search strategy, named \emph{flood-search policy}.
Algorithm~\ref{algo:search} describes the search algorithm.
In our CFG model, each node in the model represents a basic statement block. The branch edges of the branch node record the branch condition and their truth values. If the input is \texttt{G(edges, nodes)} which is a CFG model. There are two execution state lists, open and close list. At the beginning, the algorithm starts with the initialized node into the open list, while the close list is empty. Then the algorithm executes the execution states in the open list in order.  For each execution state, the algorithm will execute the shortest way from the current executed node to the exit node, and make sure the constraints in this path will be collected. In order to cover all branch edges in the graph, flood-search policy forks a copy of current execution state when it meets a fork, and adds the new execution state to the list corresponding to open or close. If all the succeeding nodes of a execution state have been visited, the execution state will be added to the close list. Comparing with other search algorithms such as breadth-first search (BFS) and depth-first search (DFS), flood-search is more suitable for dynamic symbolic execution, since flood-search in order to trigger the unvisited edges and nodes as quick as possible, meanwhile, BFS or DFS may fall in the loops.

\begin{algorithm2e}
\small
  \setcounter{AlgoLine}{0}
\caption{Flood-search for Control Flow Graph}\label{algo:search}
\DontPrintSemicolon
\newcommand\mycommfont[1]{\small\sffamily\textcolor{gray}{#1}}
\SetCommentSty{mycommfont}

\KwIn{$G(edges,nodes)$: a control flow graph}

\BlankLine

$open$ $\gets$ \{State(Root($G$))\}, $close$ $\gets$ \{\}

\Repeat{$open$.size $=$ 0}
{
 SearchShortestToExit(Pop($open$));
 \;
 \If{$open$.size $=$ 0} {
  Discharge();
 }
}

\BlankLine

\textbf{Procedure} SearchShortestToExit(State $s$)

 \If{$s$ is on the end node of the graph}{
 \Return
 }
 $next\_state$ $\gets$ Next($s$);
 \;
 \If{$s$ has unvisited edge}{
 $open \gets open \cup \{s\}$\;
 }
 \ElseIf{$s$ has visited edge}{
 $close \gets close \cup \{s\}$\;
 }
 \Else{
 $close \gets close \cup \{s\}$\;
 \Return}
 SearchShortestToExit($next\_state$);
 \;
 \;
 
\textbf{Procedure} Discharge()
 \;
 \Repeat{$close$.size $=$ 0}
 {
  $open$ $\gets$ $open$ $\cup$ \{Next(Pop($close$))\}\;
 }

\end{algorithm2e}

\subsubsection{Memory model}
The memory model is the key module to track execution states and gather constraints, by simulating whole memory allocation.
Basically, memory model stores all variables of basic types, such as \texttt{int}, \texttt{char}, including their names and values.
Dealing with complex types, such as pointers, is a challenge for analysis tools. 
Our solution is below. For the array of basic types, in addition to memory space needed by the variable in the array, the total size of the array will also be stored, to perform plus/minus on pointers, and check if the pointer is out of memory bound. Furthermore, to deal with the member of \texttt{struct}, \texttt{union} or \texttt{enum}, the total data structure and the start location in memory are recorded to locate the variable by start location and the memory offset. 

\subsection{Cloud-based Testing Service}
\begin{figure}
    \includegraphics[width=0.3\textwidth]{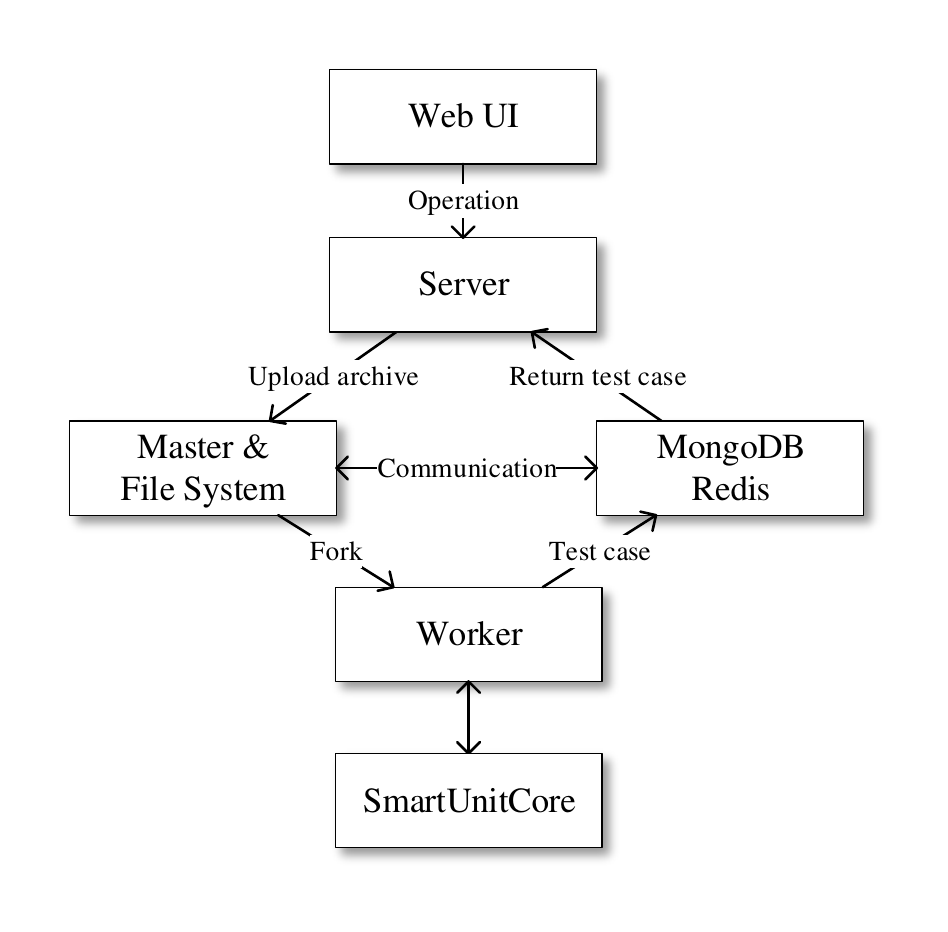}
    \caption{The workflow of the cloud-based platform.}
    \label{fig:web}
\end{figure}

The SmartUnit service contains all the other features, including web frontend UI, backend master process for handling web requests, worker process for performing analysis actions, database module for storing test results.
The SmartUnit system architecture diagram is shown in Figure~\ref{fig:web}.
The Web UI is designed to manage projects to be tested and get the results. The project under test needs to be uploaded to the server after archived, and will be passed to master process for further operations.
The master process extracts the uploaded archive, creates a record in the database, and forks a worker process to call SmartUnit analysis engine (see Figure~\ref{fig:backend}) to generate test cases for the project.

After SmartUnit finishes its analysis, worker process updates the status and saves generated test cases in the database. The status of Web UI is also updated, and then users could download test cases in specific formats, for example, \texttt{.tcf} format (for Testbed).
When the master process receives a request for a specific test case, it will check if the test case is already prepared. If not, master process will generate a test case file according to test case data in the database and the required file format. When finished, the test case is returned to the Web UI, and could be downloaded.

\subsection{Challenges and Solutions}
\subsubsection{Pointer}
The symbolizing  of the pointer variable is a challenge, because the pointer operations need to access the real value of the pointer and the variable that pointer points to. The real value of the pointer is hard to access as it is regarded as a symbolic value that means we must maintain a pointer array of pointer owner to execute a pointer operation, and the length of this array may be infinite because the pointer execution could be arbitrary. SmartUnit can support pointer operations, while a pointer memory must include the array memory of its owner which the pointer belongs to, the offset to address the position of the pointer, and some other marks such as null pointer mark. For the array of basic types, in addition to memory space needed by the variable in the array, the size of array will also be stored, to perform plus/minus operations on pointers, and check if pointer is out of memory bound.

\subsubsection{void*}
In most embedded C programs, (void*) is a special data type treated as a type that means nothing and used to transform data without data type. In Memory model, each pointer typed (void*) will be stored with its original type.
SmartUnit implements this by maintaining a void memory type and saves the alias of the memory in it. When executor comes to the (void*), it will create a \texttt{voidmemory} type, and record its type information to the aliased memory of the voidmemory. It will update the aliased memory when executing to the assignment of void*, and look up to its alias memory rather than voidmemory itself, so that we can get the type and memory information of the void* type.

\subsubsection{Complex data structure}
Complex data structures like \texttt{struct}, \texttt{union} or \texttt{enum} type are hard to handle, for they do not have a fixed length. To deal with the members of the \texttt{struct}, \texttt{union} or \texttt{enum} type, the total data structure and the start location in memory are needed to locate the variable by start location and the memory offset.
For a \texttt{struct} memory, the difficulty is to do operations related to the index. Thus, in addition to storing all variable declaration, the start and end memory location and relation between previous and next memory block are also needed to be stored.
\section{Evaluation Setup}
\subsection{Research Questions}
To evaluate our system, SmartUnit, we set up some research questions to guide our evaluation. The research questions are described as follows:
\begin{itemize}[leftmargin=*]
  \item \textbf{\textit{RQ1}}: How about the performance of automated unit test generation framework, SmartUnit, on both commercial embedded software and open-source database software?
We use statement coverage, branch coverage, and MC/DC as evaluation indicators.
  \item \textbf{\textit{RQ2}}: What factors make dynamic symbolic execution get low coverage? 
  \item \textbf{\textit{RQ3}}: Can SmartUnit find the potential runtime exceptions in real-world software?
  \item \textbf{\textit{RQ4}}: What is the difference in terms of time, cost and quality between automatically generated test cases and manually written test cases? 
\end{itemize}

We package the code under test and submit them to the SmartUnit cloud platform. SmartUnit can start dynamic symbolic execution automatically.
After the execution, a series of packaged \texttt{.tcf} files (the test cases for Testbed) for the codes can be downloaded from the platform. 
We import the test cases into Testbed to get the coverage and detect the runtime error.
Testbed will generate a detailed testing report after running the generated unit test cases.
Statement, branch and MC/DC coverage and runtime errors (\eg divided-by-zero, array index out of boundary) are provided by the report.

The statement, branch and MC/DC coverage of each function will be recorded as the performance indicator for \textit{RQ1}. 
We turn our attention to the functions which get low coverage in the three coverage criteria above, to answer \textit{RQ2}.
The runtime errors will be classified into a several of categories, so that we can obtain some insights from them, in \textit{RQ3}.
We will select some representative codes to answer \textit{RQ4}.

\subsection{Benchmark}
This paper selects two kinds of C program benchmarks for SmartUnit: commercial embedded software and open-source database software.
Table~\ref{table:benchmark} gives the list of benchmarks for the evaluation. Due to the confidentiality agreement, 
we hide the commercial software names in this paper.

\begin{table}[!t]
\footnotesize
\newcommand{\tabincell}[2]{\begin{tabular}{@{}#1@{}}#2\end{tabular}}
\renewcommand{\arraystretch}{1}
\caption{Subjects of Evaluation Benchmark Repository.}
\label{table:benchmark}
\begin{tabular}{|c||c|c|c|}
\hline
Subject &\# Files &\# Functions &\# LOC\\
\hline
\hline
aerospace software &8 &54 &3,769\\
automotive software &4 &330 &31,760\\
subway signal software &108 &874 &37,506\\
SQLite &2 &2046 &126,691\\
PostgreSQL &906 &6105 &279,809\\
\hline
\hline
Total &1028 &9,409 &479,535\\
\hline
\end{tabular}
\end{table}

\textbf{The commercial embedded software} comes from aerospace, automotive, subway signal companies.
Up to now, SmartUnit has already tested millions of code for a number of commercial embedded software. For example, 
in the aerospace company, SmartUnit has cumulatively tested more than 100,000 LOC. Over 70\% functions have achieved
more than 90\% statement coverage. 
In this paper, to conduct an intensive study, we selected three benchmarks from  different areas to ensure their diverse characteristics. All of them come from real-world industrial systems.

\textbf{The open-source database software} used in this paper are \textit{SQLite}\footnote{https://www.sqlite.org/} and \textit{PostgreSQL}\footnote{https://www.postgresql.org/}. 
Due to security demand of the commercial embedded software, we mainly use the open-source database software to explain.
We chose SQLite because it is an embedded SQL database engine, usually used in embedded software systems. 
It is a good sample for us to find some insights when using SmartUnit on the embedded system, which has nearly 130,000 LOC. 
The PostgreSQL is a representative object-relation database system, usually used as the enterprise-class database.
We chose PostgreSQL as a benchmark in order to evaluate performance and expandability of SmartUnit on the enterprise-class system.

For each subject, we put all of the \texttt{.c} and \texttt{.h} files into one folder to make it easier for Smartunit to get the dependent header files for the functions under test.
We divided each subject into an independent \texttt{.zip} package so that we can calculate coverage respectively.

\subsection{Evaluation Environment}
SmartUnit was run on a virtual machine with three processors， 3GB memory, and CentOS 7.3 operating system. Testbed (version 8.2.0) was run on a virtual machine with two processors (2.70GHz Intel(R) Core(TM) i5-2500S CPU) 1GB memory and 32bit Microsoft Windows XP Professional Service Pack 3 operating system.

\section{results and analysis}
\textbf{RQ1: How about the performance of automated unit test generation framework SmartUnit in both commercial embedded software and open-source database software?}
Table~\ref{table:RQ1_coverage} shows the coverage information of the benchmarks. In Column \emph{Subject}, the items represent the name of the programs in our benchmark. The Column \emph{\#Test cases} represents the number of test cases 
generated by SmartUnit for the corresponding benchmark.
We separated PostgreSQL into the divided modules. Thus, we used PostgreSQL plus module names as the benchmark names in Table~\ref{table:RQ1_coverage}. 
The numbers in Statement Coverage, Branch Coverage, and MC/DC Coverage represent the number of functions in the corresponding range. The number of functions which achieve 100\% coverage is highlighted in gray. \emph{N/A} means the number of functions that cannot be tested by SmartUnit or do not apply to the corresponding coverage criterion.
Our partner companies only concern those branches that have more than one conditions, when considering MC/DC coverage. Therefore, the functions that do not have branches or only have one-condition branches are counted as \emph{N/A} in MC/DC coverage.
In general, more than 90\% of functions in commercial embedded software achieve 100\% statement, branch and MC/DC coverage, 
more than 80\% of functions in SQLite and more than 60\% of functions in PostgreSQL achieve 100\% statement and branch coverage.

From the data, the conclusion is SmartUnit have a good performance on commercial embedded software and SQLite which is used in embedded systems.
The performance on PostgreSQL not as good as commercial embedded software and SQLite, but is also well enough.
It means SmartUnit is more suitable for embedded software and also have a well performance on common software.

\begin{table*}
    \Huge
    \centering
    \newcommand{\tabincell}[2]{\begin{tabular}{@{}#1@{}}#2\end{tabular}}
    \resizebox{\textwidth}{!}{\begin{minipage}{\textwidth}
    \centering
    \caption{Performance of \textit{SmartUnit} on Statement , Branch, and MC/DC coverage}
    \label{table:RQ1_coverage}
    \resizebox{\textwidth}{!}{
    \begin{tabular}{|c|c|cccccc|cccccc|cccccc|}
    \toprule
    \multicolumn{1}{|c|}{Subject} 
    &\multicolumn{1}{c|}{\#Test cases} 
    &\multicolumn{6}{c|}{Statement Coverage (\#Functions)} 
    &\multicolumn{6}{c|}{Branch Coverage (\#Functions)} 
    &\multicolumn{6}{c|}{MC/DC Coverage (\#Functions)} 
    \\

    &
    &N/A &0\%-10\% &10\%-50\% &50\%-90\% &90\%-100\% &100\% 
    &N/A &0\%-10\% &10\%-50\% &50\%-90\% &90\%-100\% &100\% 
    &N/A &0\%-10\% &10\%-50\% &50\%-90\% &90\%-100\% &100\% 

	\\ \hline
	\tabincell{c}{\emph{{aerospace software}}}
	& \tabincell{c}{368}
	& \tabincell{c}{1}
	& \tabincell{c}{-}
	& \tabincell{c}{3}
	& \tabincell{c}{6}
	& \tabincell{c}{4}
	& \tabincell{c}{\cellcolor{gray!30}41}
	& \tabincell{c}{1}
	& \tabincell{c}{-}
	& \tabincell{c}{5}
	& \tabincell{c}{5}
	& \tabincell{c}{3}
	& \tabincell{c}{\cellcolor{gray!30}41}
	& \tabincell{c}{45}
	& \tabincell{c}{2}
	& \tabincell{c}{-}
	& \tabincell{c}{-}
	& \tabincell{c}{-}
	& \tabincell{c}{\cellcolor{gray!30}8}
	
	\\ \hline
	\tabincell{c}{\emph{{automotive software}}}
	& \tabincell{c}{965}
	& \tabincell{c}{1}
	& \tabincell{c}{-}
	& \tabincell{c}{3}
	& \tabincell{c}{9}
	& \tabincell{c}{2}
	& \tabincell{c}{\cellcolor{gray!30}315}
	& \tabincell{c}{1}
	& \tabincell{c}{-}
	& \tabincell{c}{6}
	& \tabincell{c}{8}
	& \tabincell{c}{-}
	& \tabincell{c}{\cellcolor{gray!30}315}
	& \tabincell{c}{274}
	& \tabincell{c}{2}
	& \tabincell{c}{3}
	& \tabincell{c}{1}
	& \tabincell{c}{-}
	& \tabincell{c}{\cellcolor{gray!30}50}
	
	\\ \hline
	\tabincell{c}{\emph{{subway signal software}}}
	& \tabincell{c}{3617}
	& \tabincell{c}{6}
	& \tabincell{c}{-}
	& \tabincell{c}{1}
	& \tabincell{c}{24}
	& \tabincell{c}{26}
	& \tabincell{c}{\cellcolor{gray!30}817}
	& \tabincell{c}{6}
	& \tabincell{c}{-}
	& \tabincell{c}{2}
	& \tabincell{c}{29}
	& \tabincell{c}{26}
	& \tabincell{c}{\cellcolor{gray!30}811}
	& \tabincell{c}{558}
	& \tabincell{c}{6}
	& \tabincell{c}{5}
	& \tabincell{c}{11}
	& \tabincell{c}{-}
	& \tabincell{c}{\cellcolor{gray!30}294}
	
	\\ \hline
	\tabincell{c}{\emph{{SQLite}}}
	& \tabincell{c}{6945}
	& \tabincell{c}{86}
	& \tabincell{c}{6}
	& \tabincell{c}{80}
	& \tabincell{c}{147}
	& \tabincell{c}{59}
	& \tabincell{c}{\cellcolor{gray!30}1668}
	& \tabincell{c}{86}
	& \tabincell{c}{9}
	& \tabincell{c}{110}
	& \tabincell{c}{135}
	& \tabincell{c}{70}
	& \tabincell{c}{\cellcolor{gray!30}1636}
	& \tabincell{c}{1426}
	& \tabincell{c}{56}
	& \tabincell{c}{64}
	& \tabincell{c}{140}
	& \tabincell{c}{9}
	& \tabincell{c}{\cellcolor{gray!30}351}
	
	\\ \hline
	\tabincell{c}{\emph{{PostgreSQL bootstrap}}}
	& \tabincell{c}{4}
	& \tabincell{c}{18}
	& \tabincell{c}{-}
	& \tabincell{c}{1}
	& \tabincell{c}{1}
	& \tabincell{c}{-}
	& \tabincell{c}{\cellcolor{gray!30}1}
	& \tabincell{c}{18}
	& \tabincell{c}{-}
	& \tabincell{c}{2}
	& \tabincell{c}{-}
	& \tabincell{c}{-}
	& \tabincell{c}{\cellcolor{gray!30}1}
	& \tabincell{c}{19}
	& \tabincell{c}{2}
	& \tabincell{c}{-}
	& \tabincell{c}{-}
	& \tabincell{c}{-}
	& \tabincell{c}{\cellcolor{gray!30}-}
	
	\\ \hline
	\tabincell{c}{\emph{{PostgreSQL catalog}}}
	& \tabincell{c}{1023}
	& \tabincell{c}{50}
	& \tabincell{c}{2}
	& \tabincell{c}{79}
	& \tabincell{c}{117}
	& \tabincell{c}{7}
	& \tabincell{c}{\cellcolor{gray!30}170}
	& \tabincell{c}{50}
	& \tabincell{c}{6}
	& \tabincell{c}{128}
	& \tabincell{c}{69}
	& \tabincell{c}{13}
	& \tabincell{c}{\cellcolor{gray!30}159}
	& \tabincell{c}{207}
	& \tabincell{c}{151}
	& \tabincell{c}{30}
	& \tabincell{c}{8}
	& \tabincell{c}{-}
	& \tabincell{c}{\cellcolor{gray!30}29}
	
	\\ \hline
	\tabincell{c}{\emph{{PostgreSQL initdb}}}
	& \tabincell{c}{317}
	& \tabincell{c}{-}
	& \tabincell{c}{-}
	& \tabincell{c}{2}
	& \tabincell{c}{2}
	& \tabincell{c}{5}
	& \tabincell{c}{\cellcolor{gray!30}63}
	& \tabincell{c}{-}
	& \tabincell{c}{-}
	& \tabincell{c}{2}
	& \tabincell{c}{8}
	& \tabincell{c}{12}
	& \tabincell{c}{\cellcolor{gray!30}50}
	& \tabincell{c}{30}
	& \tabincell{c}{2}
	& \tabincell{c}{-}
	& \tabincell{c}{7}
	& \tabincell{c}{1}
	& \tabincell{c}{\cellcolor{gray!30}32}
	
	\\ \hline
	\tabincell{c}{\emph{{PostgreSQL pg\_dump}}}
	& \tabincell{c}{1661}
	& \tabincell{c}{14}
	& \tabincell{c}{2}
	& \tabincell{c}{25}
	& \tabincell{c}{57}
	& \tabincell{c}{22}
	& \tabincell{c}{\cellcolor{gray!30}386}
	& \tabincell{c}{14}
	& \tabincell{c}{5}
	& \tabincell{c}{34}
	& \tabincell{c}{57}
	& \tabincell{c}{37}
	& \tabincell{c}{\cellcolor{gray!30}359}
	& \tabincell{c}{345}
	& \tabincell{c}{22}
	& \tabincell{c}{21}
	& \tabincell{c}{30}
	& \tabincell{c}{2}
	& \tabincell{c}{\cellcolor{gray!30}86}
	
	\\ \hline
	\tabincell{c}{\emph{{PostgreSQL pg\_resetxlog}}}
	& \tabincell{c}{58}
	& \tabincell{c}{-}
	& \tabincell{c}{-}
	& \tabincell{c}{1}
	& \tabincell{c}{1}
	& \tabincell{c}{1}
	& \tabincell{c}{\cellcolor{gray!30}8}
	& \tabincell{c}{-}
	& \tabincell{c}{-}
	& \tabincell{c}{1}
	& \tabincell{c}{1}
	& \tabincell{c}{1}
	& \tabincell{c}{\cellcolor{gray!30}8}
	& \tabincell{c}{6}
	& \tabincell{c}{-}
	& \tabincell{c}{1}
	& \tabincell{c}{1}
	& \tabincell{c}{-}
	& \tabincell{c}{\cellcolor{gray!30}3}
	
	\\ \hline
	\tabincell{c}{\emph{{PostgreSQL pg\_rewind}}}
	& \tabincell{c}{252}
	& \tabincell{c}{4}
	& \tabincell{c}{-}
	& \tabincell{c}{1}
	& \tabincell{c}{-}
	& \tabincell{c}{6}
	& \tabincell{c}{\cellcolor{gray!30}53}
	& \tabincell{c}{4}
	& \tabincell{c}{-}
	& \tabincell{c}{1}
	& \tabincell{c}{1}
	& \tabincell{c}{6}
	& \tabincell{c}{\cellcolor{gray!30}52}
	& \tabincell{c}{48}
	& \tabincell{c}{-}
	& \tabincell{c}{-}
	& \tabincell{c}{4}
	& \tabincell{c}{-}
	& \tabincell{c}{\cellcolor{gray!30}12}
	
	\\ \hline
	\tabincell{c}{\emph{{PostgreSQL pg\_upgrade}}}
	& \tabincell{c}{312}
	& \tabincell{c}{6}
	& \tabincell{c}{-}
	& \tabincell{c}{2}
	& \tabincell{c}{5}
	& \tabincell{c}{4}
	& \tabincell{c}{\cellcolor{gray!30}83}
	& \tabincell{c}{6}
	& \tabincell{c}{-}
	& \tabincell{c}{4}
	& \tabincell{c}{3}
	& \tabincell{c}{10}
	& \tabincell{c}{\cellcolor{gray!30}77}
	& \tabincell{c}{68}
	& \tabincell{c}{2}
	& \tabincell{c}{2}
	& \tabincell{c}{9}
	& \tabincell{c}{1}
	& \tabincell{c}{\cellcolor{gray!30}18}
	
	\\ \hline
	\tabincell{c}{\emph{{PostgreSQL pg\_xlogdump}}}
	& \tabincell{c}{69}
	& \tabincell{c}{3}
	& \tabincell{c}{-}
	& \tabincell{c}{2}
	& \tabincell{c}{2}
	& \tabincell{c}{-}
	& \tabincell{c}{\cellcolor{gray!30}11}
	& \tabincell{c}{3}
	& \tabincell{c}{-}
	& \tabincell{c}{2}
	& \tabincell{c}{3}
	& \tabincell{c}{-}
	& \tabincell{c}{\cellcolor{gray!30}10}
	& \tabincell{c}{15}
	& \tabincell{c}{-}
	& \tabincell{c}{1}
	& \tabincell{c}{-}
	& \tabincell{c}{-}
	& \tabincell{c}{\cellcolor{gray!30}2}
	
	\\ \hline
	\tabincell{c}{\emph{{PostgreSQL pgtz}}}
	& \tabincell{c}{9226}
	& \tabincell{c}{589}
	& \tabincell{c}{15}
	& \tabincell{c}{588}
	& \tabincell{c}{704}
	& \tabincell{c}{69}
	& \tabincell{c}{\cellcolor{gray!30}2454}
	& \tabincell{c}{589}
	& \tabincell{c}{42}
	& \tabincell{c}{861}
	& \tabincell{c}{484}
	& \tabincell{c}{51}
	& \tabincell{c}{\cellcolor{gray!30}2392}
	& \tabincell{c}{2982}
	& \tabincell{c}{817}
	& \tabincell{c}{236}
	& \tabincell{c}{147}
	& \tabincell{c}{7}
	& \tabincell{c}{\cellcolor{gray!30}230}
	
	\\ \hline
	\tabincell{c}{\emph{{PostgreSQL psql}}}
	& \tabincell{c}{1438}
	& \tabincell{c}{3}
	& \tabincell{c}{-}
	& \tabincell{c}{12}
	& \tabincell{c}{14}
	& \tabincell{c}{16}
	& \tabincell{c}{\cellcolor{gray!30}383}
	& \tabincell{c}{3}
	& \tabincell{c}{-}
	& \tabincell{c}{14}
	& \tabincell{c}{16}
	& \tabincell{c}{19}
	& \tabincell{c}{\cellcolor{gray!30}376}
	& \tabincell{c}{336}
	& \tabincell{c}{7}
	& \tabincell{c}{12}
	& \tabincell{c}{23}
	& \tabincell{c}{3}
	& \tabincell{c}{\cellcolor{gray!30}47}
	
	\\ \hline
	\tabincell{c}{\emph{{PostgreSQL scripts1}}}
	& \tabincell{c}{197}
	& \tabincell{c}{-}
	& \tabincell{c}{-}
	& \tabincell{c}{-}
	& \tabincell{c}{7}
	& \tabincell{c}{4}
	& \tabincell{c}{\cellcolor{gray!30}30}
	& \tabincell{c}{-}
	& \tabincell{c}{-}
	& \tabincell{c}{-}
	& \tabincell{c}{7}
	& \tabincell{c}{6}
	& \tabincell{c}{\cellcolor{gray!30}28}
	& \tabincell{c}{27}
	& \tabincell{c}{-}
	& \tabincell{c}{2}
	& \tabincell{c}{6}
	& \tabincell{c}{-}
	& \tabincell{c}{\cellcolor{gray!30}6}
    \\
    
\bottomrule
\end{tabular}}
\end{minipage}}
\end{table*}

\lstset{
    frame=tb,
    language=C,
    aboveskip=3mm,
    belowskip=3mm,
    showstringspaces=false,
    columns=flexible,
    basicstyle={\small\ttfamily},
    breaklines=true,
    breakatwhitespace=true,
    tabsize=3,
    basicstyle=\footnotesize\ttfamily
}

\textbf{RQ2: What factors make dynamic symbolic execution get low coverage?}
In Table.~\ref{table:RQ1_coverage}, although SmartUnit has the high statement, branch, and MC/DC coverage,
there are some function units has a low coverage (\eg 0\%-10\%).
We found the low coverage functions, read the source code and analysis manually to find out why SmartUnit
get the low coverage in these functions. We categorized the main reasons as follows:

\textbf{Environment variable and Environment function:} 
In the benchmarks, there is a variety of environment variable and environment function in the code. 
For example, the current time is an environment variable, it comes from the system and is difficult to be symbolized.
Environment function are usually standard library calls, such as \textit{sizeof()}. 
Listing.~\ref{lst:sizeof} is an example from PostgreSQL. 
It is difficult to covert the condition in line 4 to a constraint, because symbolic execution engine is hard to comprehend the semantic of the environment functions. Therefore, the coverage can not achieve 100\% in this situation.
\begin{lstlisting}[caption={An example of environment function.},label={lst:sizeof},language=C]
static void handle_sigint(SIGNAL_ARGS)
{
        ...
        if (PQcancel(cancelConn, errbuf, sizeof(errbuf)))
        {
            CancelRequested = true;
            fprintf(stderr, _("Cancel request sent\n"));
        }
        else
            fprintf(stderr, _("Could not send cancel request: %s"), errbuf);
        ...
}
\end{lstlisting}

\textbf{Complex operation:} Although we have a solution to deal with the variable pointer, the complex operation is difficult to deal with.
Listing.~\ref{lst:complex} comes from SQLite. There is a complex pointer operation in line 4 which combines variable pointers and function pointers. 
Due to the limit of the memory model, it can not handle these complex operations.
Execution will be terminated by this kind of statement and get a low coverage.
\begin{lstlisting}[caption={An example of complex operation.},label={lst:complex},language=C]
static void callFinaliser(sqlite3 *db, int offset){
 ...
 int (*x)(sqlite3_vtab *);
 x = *(int (**)(sqlite3_vtab *))((char *)p->pModule + offset);
 if( x ) x(p);
 ...
}
\end{lstlisting}

\textbf{limitation of SMT solver:} In SmartUnit, Z3 Solver is the main SMT solver to solve the constraint. 
Although Z3 is one of the best SMT solvers, it still has some limitation. Here is an example.
\textit{y != 0 \&\& (((x-1) * y) \% y) == 1} is a constraint collected by SmartUnit in the commercial software. 
Although it is a legal constraint and there is no divided by zero faults, the Z3 solver can not deal with it.
The Z3 solver developer said that Z3 could not deal with nonlinear constraints, such as this constraint.
It is the common reason for getting the low coverage.
Thus the coverage of symbolic execution is sometimes affected by the SMT solver.

\textbf{RQ3: Can SmartUnit find the potential runtime exception in real-world software?}
From RQ1, we found that SmartUnit has a high coverage on the real world software. 
During the execution, there are also some of the potential runtime exceptions in this software.
Except the factors discussed in RQ2, we found more than 5,000 number of test cases with runtime exceptions. Due to the time limit, we have not checked every test case manually, we sampled from the test cases to analysis the runtime exceptions. Generally, we divided the runtime exceptions found by SmartUnit into three categories: 
array index out of bounds, fixed memory address and divided by zero. 

\textbf{Array index out of bounds:} As the introduction in Section 3, SmartUnit use memory model to simulate whole memory allocation.
if there is an array index out of bounds, SmartUnit will throw a runtime exception. 
Listing.~\ref{lst:array} is an example comes from SQLite. 
Obviously, in this function, there is an out of bounds runtime exception in line 10, when \emph{i \textless argc}. 
Although the caller of this function ensure \emph{i \textless= argc} in SQLite, 
it has a potential runtime exceptions if other callers not ensure \emph{i \textless= argc}.
There is even no precondition in the comment.
It is quite serious if a programmer wants to call it, but doesn't know the precondition.
We have found much of runtime exceptions in this category from all of the benchmarks.

\begin{lstlisting}[caption={An example of array index out of bounds.},label={lst:array},language=C]
/*
** Get the argument to an --option.  Throw an error and die if 
** no argument is available.
*/
static char *cmdline_option_value(int argc, char **argv, int i) {
    if (i == argc) {
        utf8_printf(stderr, "%s: Error: missing argument to %s\n", argv[0], argv[argc - 1]);
        exit(1);
    }
    return argv[i];
}
\end{lstlisting}

\textbf{Fixed memory address:} Fixed memory address is a problem in dealing with the variable pointer. 
In the embedded system, there are many pointer operations with the fixed memory address. 
It is hard for memory model to simulate a fixed memory address.
For example, the operations like (*0X00000052) or (* (symbolic variable + 12)) will cause runtime exceptions.
It usually gets \textit{NULL} when referring from a fixed memory address.
Thus symbolic execution will throw runtime exception with fixed memory address operations.

\textbf{Divided by zero:} Divided by zero is also a common runtime exception in the benchmarks. 
It usually appears in numerical calculation of program. 
SmartUnit will generate boundary value to check if the program exists divided by zero runtime exception.
We select a brief example from SQLite to discuss this runtime exception. 
Listing.~\ref{lst:zero} is the function which has the potential runtime exception in line 19.
When \emph{nUsable == 4}, the expression \emph{(nTotal - nMinLocal) \% (nUsable - 4)} will throw a divided by zero runtime exception.
we read the source code manually and found that it is difficult to comfirm whether \emph{nUsable} could be 4 or not.
The comment of the function does not contain the precondition of \emph{nUsable}.
It's horrible if the programmer who calls this function in a new function and do not know the implicit prediction.

\begin{lstlisting}[caption={An example of divided by zero.},label={lst:zero},language=C]
static void getLocalPayload(
  int nUsable,                    /* Usable bytes per page */
  u8 flags,                       /* Page flags */
  int nTotal,                     /* Total record (payload) size */
  int *pnLocal                    /* OUT: Bytes stored locally */
){
  int nLocal;
  int nMinLocal;
  int nMaxLocal;
 
  if( flags==0x0D ){              /* Table leaf node */
    nMinLocal = (nUsable - 12) * 32 / 255 - 23;
    nMaxLocal = nUsable - 35;
  }else{                          /* Index interior and leaf nodes */
    nMinLocal = (nUsable - 12) * 32 / 255 - 23;
    nMaxLocal = (nUsable - 12) * 64 / 255 - 23;
  }

  nLocal = nMinLocal + (nTotal - nMinLocal) % (nUsable - 4);
  if( nLocal>nMaxLocal ) nLocal = nMinLocal;
  *pnLocal = nLocal;
}
\end{lstlisting}

In summary, SmartUnit could find the potential runtime exceptions in real-world software. We categorized and analyzed the potential runtime exceptions. 
Most of these potential runtime exceptions existed, due to there is no protection for input values.
At the same time, there is no clear precondition specification for the functions.
Although there are protection codes in the caller of the function, the potential runtime exceptions may cause the real faults in real-world software.

\textbf{RQ4: What is the difference between automatically generated test cases and manually written test cases?}
In RQ4, we compare automatically generated test cases and manually written test cases in the following aspect: time, cost and quality.

\textbf{Time \& Cost:} Table~\ref{table:RQ4_time} shows the test set generated time for each benchmark. 
The column \textit{Subject} list all of the benchmark we used.
As Table~\ref{table:RQ1_coverage}, we separate PostgreSQL into the individual module, and name with PostgreSQL plus module name.
The second column \textit{\# Functions} represents the number of functions in the benchmark.
The column \textit{Time} means the test set generated time for the corresponding benchmark.
The column \textit{Average} represents the average test set generated time for the corresponding benchmark per function, 
in other words, it means the average time of SmartUnit generates test set for a function.
Deserve to be mentioned, we use \textit{Total\# Function} and \textit{Total Time} to calculate \textit{Total Average}.

In Table~\ref{table:RQ4_time}, it is obvious that the average time ranges from 1s to 7s and total average time is 3.77s for all of the benchmarks.
It means that SmartUnit spends nearly 4s to generate test cases for a function, 
and in the best situation, more than 90\% of the functions can achieve 100\% statement, branch, and MC/DC coverage.
How about the test engineer? In our survey from companies, a trained test engineer can product test case for 5-8 functions per day. 
Thus using automated unit test generation framework can cost less time than employing test engineers. 

The main cost of automatically generated test cases and manually written test cases are both the salary paid for the employee. We assume a
developer costs twice salary higher than a test engineer. SmartUnit has cost 24 man-month to release. For manually testing, 
a test engineer could write the test case for about 160 functions per month. Thus 24 man-month cost could support test engineers write about 8,000 functions. 
In summary, if you have a large number of function units (\eg more than 10,000), automatically generated test cases are cheaper.
On the contrary, manual test cases are cheaper for the little scale of projects. 

\textbf{Quality:} From RQ1, we have the conclusion that more than 90\% of functions in commercial embedded software can achieve 100\% statement, branch and MC/DC coverage; more than 80\% of functions in SQLite and more than 60\% of functions in PostgreSQL achieve 100\% statement and branch coverage. According to our survey in the companies, the test engineers need to achieve 100\% coverage for each function. If we use coverage as a quality indicator, automatically generated test case has overwhelming superiority on cost and time though manual test case has higher coverage in some cases. Meanwhile, the automatically generated test cases could find runtime exception in time. 

\begin{table}[!t]
\footnotesize
\newcommand{\tabincell}[2]{\begin{tabular}{@{}#1@{}}#2\end{tabular}}
\renewcommand{\arraystretch}{1}
\caption{Test set generated time for each benckmark.}
\label{table:RQ4_time}
\begin{tabular}{|c||c|c|c|}
\hline
Subject &\# Functions &Time (s) &Average (s/func) \\
\hline
\hline
aerospace software &54 &318 &6\\
automotive software &330 &329 &1\\
subway signal software &874 &2,476 &3\\
SQLite &2046 &13,482 &6\\
PostgreSQL bootstrap &21 &48 &2\\
PostgreSQL catalog &425 &1,350 &3\\
PostgreSQL initdb &72 &548 &7\\
PostgreSQL pg\_dump &506 &3,428 &7\\
PostgreSQL pg\_resetxlog &11 &71 &6\\
PostgreSQL pg\_rewind &64 &352 &5\\
PostgreSQL pg\_upgrade &100 &465 &5\\
PostgreSQL pg\_xlogdump &18 &130 &7\\
PostgreSQL pgtz &4419 &10,478 &2\\
PostgreSQL psql &428 &1,676 &4\\
PostgreSQL scripts1 &41 &311 &7\\

\hline
\hline
Total &9,409 &35,462 &3.77\\
\hline
\end{tabular}
\end{table}

\textbf{Discussion}

Traditional automated unit testing tools focus on the automated random testing, \eg, test cases are randomly derived as the input to invoke the program code under test. Although the random test data generation algorithm is quite easy to be implemented, it is obviously that the expected coverage criteria (e.g., the branch and the statement coverage) cannot be guaranteed. The DART (Directed Automated Random Testing) is a classical unit testing tool that supports the directed random test data generation~\cite{godefroid2005dart}. One unique characteristic of the DART is its complete automation for unit testing. Since its underlying approach is random testing, the DART focuses on detecting standard errors such as program crashes, assertion violations, and non-termination while it is not deliberately designed for certain coverage criteria requested by the industrial standards. Our SmartUnit framework can handle the automated testing and support various coverage criteria of the industrial standards.

   PEX is a famous unit testing tool that automatically generates test suites with high code coverage~\cite{tillmann2008pex}. It is a Visual Studio add-in for testing .NET Framework applications. In practice, the PEX tool is adapted to testing the C\# code. Similarly the IntelliTest\footnote{https://msdn.microsoft.com/en-us/library/dn823749.aspx} also automatically generates test data for the unit test of C\#. The basic idea of the IntelliTest is to generate test data to execute each statement of the target code and then analyze the coverage of the conditional branch. The Microsoft Security Risk Detection is a unique fuzz testing service for finding security critical bugs in software. These three tools significantly improve the unit testing in bug detection and time efficiency. However, since these tools are designed as general solutions for unit testing, it cannot be directly applied in the domain of real-time control software testing. They fall short in supporting the unit testing of the control software since some unique features such as the high coverage request of particular criteria requested by the industrial standards have not been deliberately considered. The SmartUnit framework focuses on the request of current industry standards and offers a completely automated solution for the unit testing of the real-time software.
   
Although there is a plug-in of Testbed called LDRA TBrun\footnote{http://ldra.com/industrial-energy/products/tbrun/} which can automatically generate driver program and test harness without manual script,
SmartUnit is quite different from it.
LDRA TBrun uses data dictionary to generate the test cases, it combines several values in the data dictionary, 
usually considering up bound and low bound of variables. 
This strategy is not able to deal with memory calculation, while SmartUnit can easily catch memory change constraints by DSE.
LDRA TBrun generates unit test suite based-on boundary value (\eg, maximum value, minimum value, median value, \etc),
while SmartUnit is a coverage-driven tool, it can satisfy statement, branch, MC/DC coverage criteria from industrial requirements.
SmartUnit can avoid repetitive test cases, while LDRA TBrun has many repetitive test cases in its test set.
LDRA TBrun can not give the expectative output value for each test case, 
but SmartUnit can expect output value by executing test cases automatically after generation.
SmartUnit can not only generate the test cases for function parameters and global variables which support in LDRA TBrun,
it can also generate the test cases for instrument function parameters and instrument function return value. 
In summary, LDRA TBrun is suitable for critical testing and robustness testing,
while SmartUnit is suitable for coverage-driven testing.
\section{related work}

This section discusses related work in two aspects: symbolic execution and automated unit test generation.

Symbolic execution is a classic software testing technique, 
and recently enhanced with the dynamic symbolic execution (also called concolic testing) technique~\cite{cadar2013symbolic,cadar2011symbolic,sen2006scalable}.
There are several symbolic execution tool, such as Pex~\cite{tillmann2008pex} for .NET, 
Java PathFinder~\cite{anand2007jpf}, jCUTE~\cite{sen2006cute} for Java,
KLEE~\cite{cadar2008klee,cadar2008exe}, DART~\cite{godefroid2005dart}, CAUT~\cite{su2014automated}, CUTE~\cite{sen2005cute,sen2006cute}, CREST~\cite{burnim2008heuristics} for C.

There is also much work on automated test generation. Some of them are based on Java.
RANDOOP~\cite{pacheco2007randoop}, EVOSUITE~\cite{fraser2011evosuite}, AGITARONE\footnote{http://www.agitar.com/solutions/products/agitarone.html} are usually used to generate test cases and evaluate on real-world software~\cite{shamshiri2015automatically, fraser2014large, almasi2017industrial}. 
Some of them focus on unit test generation~\cite{robinson2011scaling, zheng2010random, garg2013feedback, xie2003tool, zhang2011combined}. 
SmartUnit has a good performance on unit test generation, but also is evaluated on real-world software and used in practice. In the future, we will implement more advanced coverage criteria, \eg, data-flow coverage~\cite{Su17}, to further improve its fault detection ability; and extend SmartUnit to support other types of software (\eg, mobile applications~\cite{SuAppTesting}) and scenarios (\eg, requirement testing~\citep{MiaoPYSB0CX16}).

In industry, \textit{Microsoft} developed \textit{Unit Meister} for parameterized unit tests~\cite{tillmann2005parameterized} and \textit{SAGE} for whitebox fuzzer testing~\cite{godefroid2012sage}. 
\textit{Fujitsu} tried to use symbolic execution to generate test~\cite{tokumoto2012enhancing}, while \textit{Samsung} used \textit{CREST} and \textit{KLEE} on mobile platform programs~\cite{kim2011concolic,kim2012industrial}.
The benchmarks they chose were all from their software. While We chose the benchmarks from a variety of embedded software systems and open-source software.

\section{conclusion}

In this paper, we propose an automated unit coverage-based testing tool for embedded software called SmartUnit.
It comprises of a dynamic symbolic execution engine, a unit test generator, and a private cloud-based service.
It has been used in a series of real-world embedded software projects such as aerospace, airborne and ground-based systems.
This tool has been developed and improved collaboratively with several top companies in China.
The companies have used the SmartUnit in their daily testing process and improved their software reliability.
We show a general pattern of how to use symbolic execution in practice with the example of SmartUnit.

The performance of SmartUnit is evaluated by testing with both commercial embedded software and open-source software. Besides, some runtime exceptions detected by our tool are collected and classified as guidance for potential users to avoid such runtime exceptions in software developing process. Challenges in using dynamic symbolic execution in the industrial environment have also been discussed.

In summary, although there are some challenges, it is possible to use dynamic symbolic execution technique on real-world software and get a high performance on coverage criteria. It is also practicable to build automated unit test generation tool as a cloud service to make unit testing easier to be adopted. 

\begin{acks}
We would like to thank the anonymous reviewers for their valuable feedback. Ting Su is partially supported by NSFC Projects No. 61572197 and No. 61632005. Geguang Pu is partially supported by MOST NKTSP Project 2015BAG19B02 and STCSM Project No. 16DZ1100600.
Chengyu Zhang is partially supported by  China HGJ Project (No. 2017ZX01038102-002).

\end{acks}

\bibliographystyle{ACM-Reference-Format}
\bibliography{sample-bibliography} 

\end{document}